\documentclass{desyproc}

\begin{document}
\title{CP-Conservation in QCD and why\\
 only ``invisible'' Axions work}

\author{{\slshape  Jihn E. Kim}\\[1ex]
Center for Axion and Precision Physics (IBS), 291 Daehakro, Daejeon 34141,  Korea}

\contribID{familyname\_firstname}

\confID{13889}  
\desyproc{DESY-PROC-2016-XX}
\acronym{Patras 2016} 
\doi  

\maketitle

\begin{abstract}
Among solutions of the strong CP problem, the ``invisible'' axion  in the narrow axion window is argued to be the remaining possibility among natural solutions on the smallness of $\bar{\theta}$. Related to the gravity spoil of global symmetries, some prospective invisible axions from theory point of view are discussed. In all these discussions, including the observational possibility, cosmological constraints must be included. 
\end{abstract}

\newcommand{\dis}[1]{\begin{equation}\begin{split}#1\end{split}\end{equation}}
\newcommand{\ie}{{\it i.e.~}}
\newcommand{\etal}{{\it et al.\,}}

\newcommand{\cred}[1]{{\color{red}{#1}}}
\newcommand{\cblue}[1]{{\color{blue}{#1}}}
\newcommand{\cgreen}[1]{{\color{green}{#1}}}

 \newcommand{\dell}{\delta_{\rm PMNS}}
\newcommand{\delq}{\delta_{\rm CKM}}
\newcommand{\delB}{\delta_{\rm B}}
\newcommand{\delL}{\delta_{\rm L}}
\newcommand{\delx}{\delta_{\rm X}}
\newcommand{\eL}{\epsilon_{\rm L}}
 
\newcommand{\Qem}{Q_{\rm em}}
\newcommand{\QPQ}{Q_{\Gamma}}
\newcommand{\cagg}{c_{a\gamma\gamma}}
\newcommand{\caggb}
{\tilde{c}_{a\gamma\gamma}}
\newcommand{\tev}{\,\textrm{TeV}}
\newcommand{\gev}{\,\textrm{GeV}}
\newcommand{\meV}{\,\mathrm{MeV}}
\newcommand{\keV}{\,\mathrm{keV}}
\newcommand{\eV}{\,\mathrm{eV}}
\newcommand{\Mp}{M_{\rm P}}
\newcommand{\Mpt}{$M_{\rm P}$}
\newcommand{\Mg}{{M_{\rm GUT}}}
\newcommand{\Mgt}{$M_{\rm GUT}$} 
\newcommand{\vew}{$v_{\rm ew}$} 
\newcommand{\axino}{\tilde{a}}
\newcommand{\NDW}{$N_{\rm DW}$}

\newcommand{\Uone}{U(1)$_{\rm gl}$}
\newcommand{\Uga}{U(1)$_{\rm ga}$}
\newcommand{\UPQ}{U(1)$_{\rm PQ}$}
\newcommand{\UR}{U(1)$_{\rm R}$}
\newcommand{\UG}{U(1)$_{\Gamma}$}
 
\def\sw0{{$\sin^2\theta_W^0$}}
\def\thetab{$\bar\theta$}
\def\theq{$\theta_{QCD}$}
\def\thew{$\theta_{weak}$}

\def\Tr{{\rm Tr \,}}
\def\Nr{{\rm N}}
\def\Nc{{\cal N}}
\def\NDW{{N}_{\rm DW}}
\newcommand{\Z}{{\bf Z}}

\def\smg{{SU(3)$_C\times$SU(2)$_W\times$U(1)$_Y$}}
\def\E6{{\rm E_6}}

\def\EE8{{\rm E_8\times E_8'}}
\def\GG{SU$(5)_{\rm GG}$}
\def\flip{SU$(5)_{\rm flip}$}
\def\anti{anti-SU$(5)$}
\def\antiFD{SU$(5)_{\rm anti2}$}
\def\antiSD{SU$(7)_{\rm anti2}$}
\def\antiED{SU$(8)_{\rm anti2}$}
\def\antinD{SU$(N)_{\rm anti2}$}
\def\antinT{SU$(N)_{\rm anti3}$}
\def\antinQ{SU$(N)_{\rm anti4}$}
\def\fourb{\overline{\bf 4}\,}
\def\four{{\bf 4}}
\def\six{{\bf 6}}
\def\threeb{\overline{\bf 3}\,}
\def\three{{\bf 3}}
\def\fif{{\bf 15}}
\def\sixb{\overline{\bf 6}\,}
\def\six{{\bf 6}}

\def\one{\bf 1}
\def\oneb{\bar{\bf 1}}
\def\two{\bf 2}
\def\five{\bf 5}
\def\ten{\bf 10}
\def\tenb{\overline{\bf 10}}
\def\fsix{{\bf 56}} 

\def\fiveb{\overline{\bf 5}}
\def\threeb{{\bf\overline{3}}}
\def\three{{\bf 3}}
\def\ts{{\bf 27}}
\def\tsb{$\overline{\bf 27}$}
\def\fb{{\overline{F}\,}}
\def\hb{{\overline{h}}}
\def\Hb{{\overline{H}\,}}

\section{Introduction}

From a fundamental theory point of view,  presumably the most fundamental parameters are given at a mass-defining scale which is considered to be the Planck mass $\Mp\simeq 2.43\times 10^{18\,}\gev$.  In this sense, we consider that a natural mass scale is the Planck mass and/or some scale  suppressed by a small coupling compared to $\Mp$, i.e. such as the GUT scale.

Then, all smaller mass scales much smaller than $\Mp$ are better to come from some symmetry arguments. Such symmetries are chiral symmetry for  fermions and global symmetries for scalars.   The smallness of electron mass compared to $\Mp$ by a factor of $10^{-22}$ is better come from a chiral symmertry. In the standard model(SM), the chiral symmetry is intrinsically implemented in the fermion representations in the SM. Even though the original Kaluza-Klein model for electromagnetism looked impressive, it fails badly here \cite{Witten83}. A spontaneously broken chiral symmetry leads to a massless Goldstone boson \cite{Kibble67}.
Among Goldstone bosons, “invisible” axion $a$ \cite{InvAxions} is the most interesting one as this conference witnesses. The  “invisible”  axion arises from spontaneously breaking the Peccei-Quinn(PQ) global symmetry \cite{PQ77}.

\section{The strong CP problem}

Because of instanton solutions of QCD, there exists a gauge invariant vacuum parametrized by an angle $\bar\theta$ which is the coefficient of  the gluon anomaly,  $G^a\tilde{G}^a$. It is the flavor singlet and acted as the source for solving the U(1) problem of QCD \cite{Hooft86}.
Thus, this $\bar\theta$ term is physical, but it leads to the so-called
``strong CP problem'': it comes from the smallness of neutron electric dipole moment(nEDM), ``Why is the nEDM so small?'' \cite{InvAxions}.
In this regard, we try to explain why the “invisible” axion is the remaining solution in natural schemes toward understanding the strong CP problem as presented in the title.
\begin{figure}[t!]
\centerline{\includegraphics[width=0.7\textwidth]{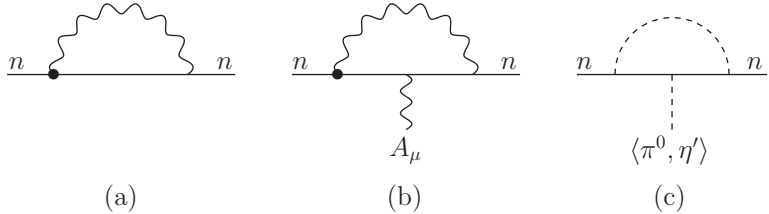}}
\caption{The bullets in (a,b) are the insertions of CP violating interaction shown in (c). VEVs of $\pi^0$ and $\eta'$ break CP. (b) leads to nEDM.}\label{Fig:nEDM}
\end{figure}

If the CP violating coupling $\overline{g_{\pi NN}}$ is present, nEDM is calculated to be (with the CP conserving ${g_{\pi NN}}$ term)
\dis{
\frac{d_n}{e}=\frac{{g_{\pi NN}}\overline{g_{\pi NN}}}{4\pi^2 m_N} \ln\left(\frac{m_N}{m_\pi}\right).\label{eq':nEDM}
}
Figure \ref{Fig:nEDM}\,(c) shows $\overline{g_{\pi NN}}$, and Figure \ref{Fig:nEDM}\,(b) shows  Eq. (\ref{eq':nEDM})  \cite{Crewther79}.
If the $\bar\theta$ term is present in QCD, then $\pi^0$ can obtain a VEV and $|\overline{g_{\pi NN}}|=\bar\theta/3$ \cite{InvAxions}. The non-observation of nEDM put a limit on $|\bar\theta|$ as less than $10^{-10}$.

The most widely discussed solutions in the SM are (i) the ``invisible'' axion, (ii) massless up-quark, and (iii) calculable  $\bar\theta$ models. The massless up-quark possibility is ruled out phenomenologically, $m_u=2.15(0.15)\,$MeV \cite{PDGqmass}. The calculable models start with CP invariant Lagrangian, and calculate loop effects that the results are safe enough such that the loop generated   $\bar\theta$ is less than  $10^{-10}$.
Thus, calculable models within the SM gauge group, \smg, are usually considered: the so-called Nelson-Barr type models \cite{Nelson,Barr}. But, it is usually very difficult to remove the $\bar\theta$ term up to two-loop level. Barring the possibility of gauge or global symmetries, one introduces some kind of discrete symmetry toward this objective \cite{Segre79}, or some structure on the coupling texture \cite{Barr}. The texture possibility may rely on discrete symmetries. Spontaneously broken discrete symmetries lead to cosmological domain wall problem. Thus, the calculable possibility is not so attractive compared to introducing just one spontaneously broken global symmetry which we discuss below.

\section{The PQ symmetry}

Since 1964, there has been always a need for a theory of weak CP violation. The strong CP problem is interwined with the weak CP, $\bar{\theta}=\theta_{\rm QCD}+\theta_{\rm weak}$, where $\theta_{\rm weak}$ is the one contributed by the weak CP violation. Now, the  Kobayashi-Maskawa(KM) model \cite{KM73} is accepted for the SM  weak CP violation, and there is no need to consider the weak CP in gauge groups beyond \smg. But, in the middle of 1970's, it was not so and many introduced their own models of weak CP violation and even in \smg ~there were many \cite{Mohapatra72,KM73,Weinberg76,LeeBW77,Georgi78}. For example, the first gauge theory model was Mohapatra's \cite{Mohapatra72}, which appeared as the 2nd example of Ref. \cite{KM73} but discarded there with the statement `phenomenologically unacceptable'. One notable one is Weinberg's weak CP violation model \cite{Weinberg76} which was presented when he  used the sabbatical year at SLAC. It was the time when the third quark family was not discovered (even though $\tau$ was discovered), and he tried to introduce the weak CP violation in the Higgs potential, without introducing the third family quarks. With two Higgs doublets, he applied the Glashow-Weinberg method not to have tree-level flavor-changing neutral processes, \ie $H_u$ coupling to up-type quarks and  $H_d$ coupling to down-type quarks  \cite{GW77}. Then, the weak CP violation mattered in the potential,
\dis{
V_{\rm W}=\frac12\sum_I m_I^2\phi_I^\dagger\phi_I+\frac14 \sum_{IJ} \left\{a_{IJ} \phi_I^\dagger\phi_I\phi_J^\dagger\phi_J+b_{IJ}\phi_I^\dagger\phi_I\phi_J^\dagger\phi_J +(c_{IJ} \phi_I^\dagger\phi_I\phi_J^\dagger\phi_J+{\rm H.c.})
\right\}\label{eq:Weinberg}
}
Weinberg's necessary condition for the existence of CP violation is non-zero $c_{IJ}$ terms. Peccei and Quinn noticed that there emerges a global symmetry by removing  the $c_{IJ}$ terms in $V_{\rm W}$, and has the famous PQ potential
 \dis{
V_{\rm PQ}=\frac12\sum_I m_I^2\phi_I^\dagger\phi_I+\frac14 \sum_{IJ} \left\{a_{IJ} \phi_I^\dagger\phi_I\phi_J^\dagger\phi_J+b_{IJ}\phi_I^\dagger\phi_I\phi_J^\dagger\phi_J \right\}\label{eq:PQ}
}
which is the basis for the Peccei-Quinn-Weinberg-Wilczek(PQWW) axion \cite{PQWW77}. Soon, it was ruled out chiefly by the results on beam dump experiments \cite{Peccei78}.  Non-observation of the PQWW axion triggered interests in calculable models after whose period the ``invisible'' axion replaced the PQWW axion in the solutions of the strong CP problem. The PQWW axion is not relevant in this conference which will discuss mostly on the cosmic axion search.
 
\section{The ``invisible'' axion}
After the above tries, finally the ``invisible'' axions are discovered. These are known as the KSVZ axion \cite{KSVZ1,KSVZ2} and the DFSZ axion \cite{DFSZ}.   As we witnessed in the previous section, if only a few terms in all possible terms in $V$ are considered, there may appear global symmetries.
\begin{figure}[t!]
\hskip 1cm {\includegraphics[width=0.37\textwidth]{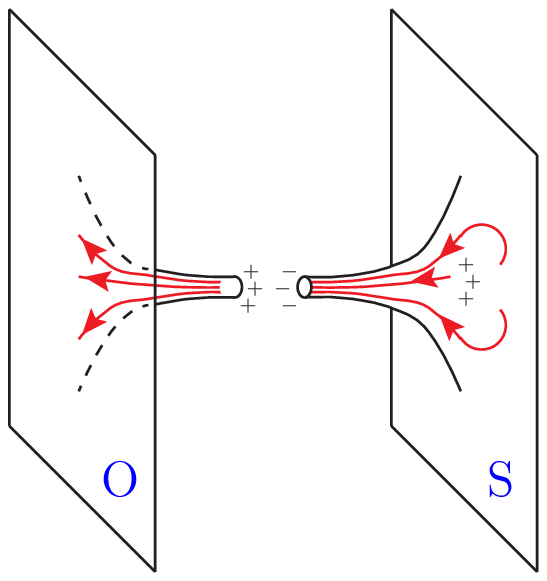}}\hskip 1cm {\includegraphics[width=0.4\textwidth]{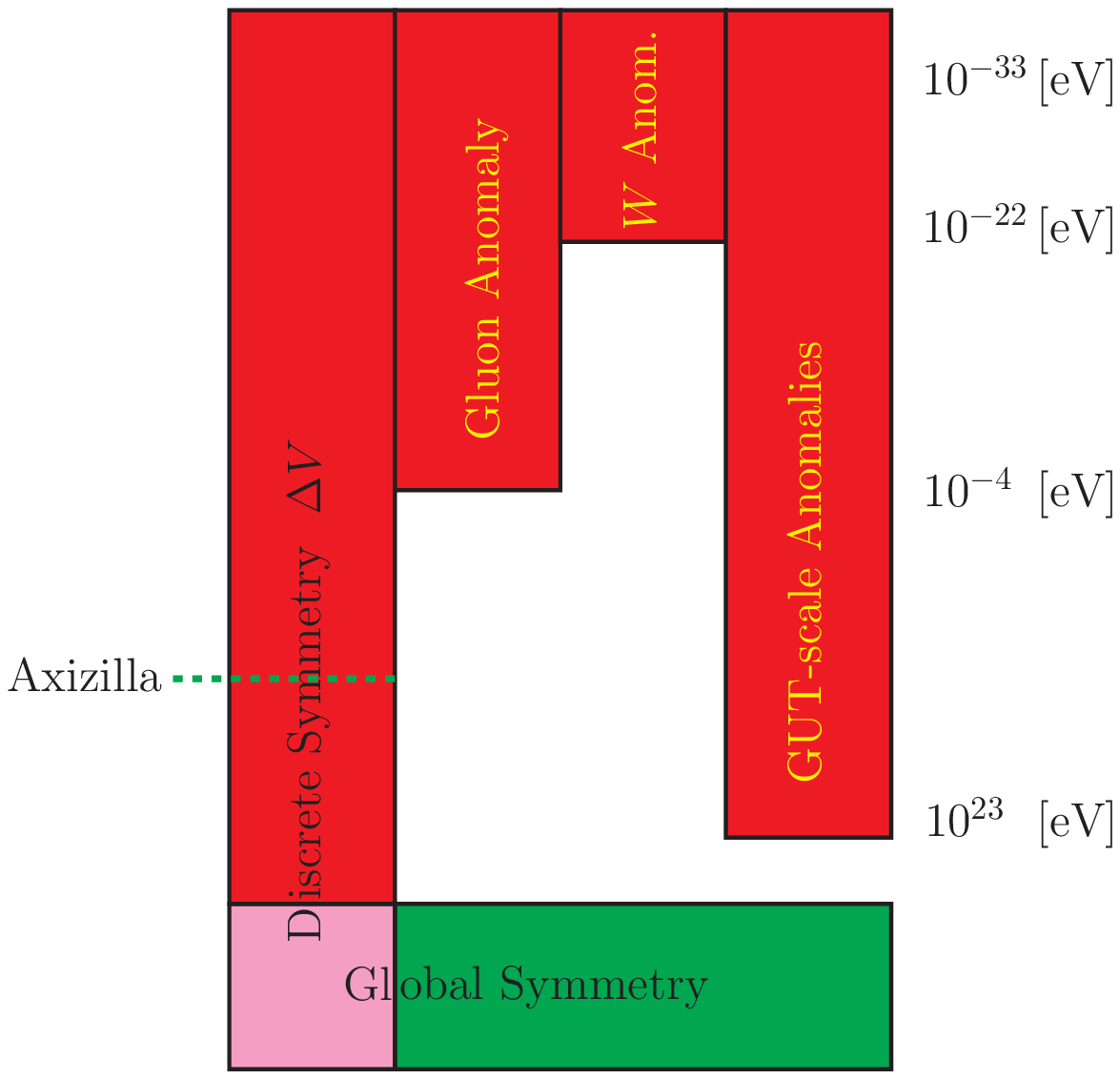}}\vskip 1cm
\centerline{(a)\hskip 6cm (b)}
\caption{(a) A wormhole connecting our world to a shadow world, and (b) terrms allowed by discrete and global symmetries.}\label{Fig:Discrete}
\label{sec:Discrete}
\end{figure}
It has a profound implication in some ultra-violet completed theories. It is known that quantum gravity does not allow global symmetries. However, 
gauge symmetries are allowed. One possible quantum gravity phenomenon, the wormhole, can connect our Universe to a shadow world.  Any method to obtain an effective interactiion by cutting off the wormhole will send back to our Universe the flown-out gauge charges due to the existence of flux lines. Figure \ref{sec:Discrete}\,(a) depicts this phenomenon.  
\begin{wraptable}{r}{0.45\textwidth}
\centerline{\begin{tabular}{|r|c|} 
\hline &\\[-1em]
KSVZ:  $Q_{\rm em}$ ~&$c_{a\gamma\gamma}$\\\hline
$0$ ~~ &$-2$~ \\[0.3em]
$\pm \frac13$~~  &$-\frac{4}3$~  \\[0.3em]
$\pm \frac23$~~  &$ \frac{2}3$   \\[0.3em]
$\pm 1 $~~  &$4$  \\[0.3em]
$(m,m)$~&$-\frac{1}3$~  \\[0.2em] 
\hline
\end{tabular}}
\caption{KSVZ model with $m_u=0.5\, m_d$. $(m,m)$ in the last row means $m$ quarks of $Q_{\rm em}=\frac23\,e$ and  $m$ quarks of $Q_{\rm em}=-\frac13\,e$.}
\label{tab:KSVZ}
\end{wraptable}
But, global charges do not carry flux lines and it is considred that global symmetries are not respected by gravity. 
Figure \ref{sec:Discrete}\,(b) shows all allowed couplings if some discrete symmetry is present. However, if we consider only a few dominant terms shown in the lavender square, then there can appear some global symmetries.

\begin{wraptable}{r}{0.45\textwidth}
\centerline{\begin{tabular}{|r c|c|} 
\hline &&\\[-1em]
DFSZ:  $(q^c$-$e_L)$ pair ~& Higgs &$c_{a\gamma\gamma}$\\\hline\\[-1.2em]
&&  \\[-0.85em]
non-SUSY $(d^c,e)$ ~~ & $H_d$&$\frac23$~ \\[0.3em]
non-SUSY $(u^c,e)$ ~~  & $H_u^*$&$-\frac43$~ 
  \\[0.2em] 
GUTs  ~~ &  &$\frac23$~ \\[0.3em]
SUSY   ~~  &  &$\frac23$~ 
  \\[0.2em] 
\hline
\end{tabular}}
\caption{DFSZ model with $m_u=0.5\, m_d$.}
\label{tab:DFSZ}
\end{wraptable}

  These global symmetries are approximate and broken by terms in the red parts. The red part adjacent to the lavender symbolizes the terms in $V$. One such example is heavy axions or axizilla \cite{KimAxizilla}, which cannot solve the strong CP problem. 
The reds not connected to the lavender break the global symmetry by non-Abelian gauge anomalies, and the QCD axion obtains mass basically by this term. In any case, a PQ global symmetry can be obtained at least approximately from an ultra-violet completed theory. 

  Initially, ``invisible'' axions were  ignored as ``invisible'', but there surfaced a dim hope of detecting it by cavity detectors \cite{Sikivie83}. Then, there was a need to calculate the axion-photon-photon coupling, which has been done in Ref. \cite{Kim98,InvAxions} for several invisible axion models. So far, all the existing calculations have been published by the author's group alone \cite{Kimjhep07,Kimplb15,KimNam16} with some confirmation by J. Ashfaque, H.P. Nilles and P. Vaudrevange.
 
 Table \ref{tab:KSVZ} shows the axion-photon-photon couplings in several KSVZ models. Table \ref{tab:DFSZ} shows  the axion-photon-photon couplings in several DFSZ models. Here, $H_d$ and $H_u^*$ imply that they give mass to $e$. GUTs and SUSY choose appropriate Higgs doublets and always give $c_{a\gamma\gamma}=\frac23$.
   
\begin{wraptable}{r}{0.45\textwidth}
\centerline{\begin{tabular}{|l| r|c|} 
\hline  &&\\[-1em]
String:  &$c_{a\gamma\gamma}$  &Comments\\\hline\\[-1.2em]
&&  \\[-0.9em]
Ref. \cite{Kimjhep07}  &  $-\frac13$~ &Approximate\\[0.3em] 
Ref. \cite{Kimplb15,KimNam16} &  $\frac23$~ &Anom. U(1)\\[0.3em]
\hline
\end{tabular}}
\caption{String model. Comments are for \UPQ. In the last line, $c_{a\gamma\gamma}= (1-2\sin^2\theta_W)/\sin^2\theta_W$ with $m_u=0.5\, m_d$.}
\label{tab:String}
\end{wraptable}
   
Table \ref{tab:String} shows $c_{a\gamma\gamma}$ for a few string compactifications \cite{KimKyaeZ12,HKK09} from heterotic string. In these string compactifications, it is required to start from a model allowing correct SM phenomenologies. This is the reason that there has not appeared any calculation from intersecting brane models.
 
Therefore, it is a key question how the PQ symmetry is defined. As we witnessed in Eqs. (\ref{eq:Weinberg}) and (\ref{eq:PQ}), a small number of terms in $V$ may have a chance to introduce  a global symmetry. However, the way Eq. (\ref{eq:PQ}) was declared is ad hoc. It amounts to declaring a global symmetry from the outset. It is better if some ultra-violet theory forbids some terms such as $c_{IJ}$ of Eq. (\ref{eq:Weinberg}). The well-known effective interacion example allowed by gauge symmetry is Weinberg's neutrino mass in the SM,  $\ell\ell H_u H_u/M$ where $M$ is the scale where a more satisfactory theory is defined. In this vein, suppose the phase of the SM singlet $\sigma$ is ``invisible'' axion \cite{KSVZ1}. Then, the PQ symmetry defining term can be a renormalizable term, $H_uH_d\,\sigma^2$. But, then one must fine-tune the coefficient of this term to have a small ratio $v_{\rm ew}/f_a$, which is the reason that the non-SUSY DFSZ model has a fine-tuning problem \cite{Dreiner14}. For the KSVZ model, the intermediate scale interaction is
\dis{
\bar{Q}\,Q\,\sigma+{\rm H.c.}
}
where $Q$ is a heavy quark,
and the VEV of $\sigma$ is determined by parameters at the intermediate scale. At the electroweak scale, its effect appears as $(a/f_a)G^a_{\mu\nu}\tilde{G}^{a\,\mu\nu}$ and no fine-tuning is needed even  in this non-SUSY SM. In SUSY models, one may consider a renormalizable superpotential term 
$H_uH_d\,\sigma$ for defining the PQ symmetry. Then, a natural scale for the VEV of $\sigma$ is  $v_{\rm ew}$, which is ruled out phenomenologically. We can use some discrete symmetry to forbid it as depicted in Fig. \ref{Fig:Discrete}\,(b). Then, the most important term in the superpotential defining the PQ symmetry is $H_uH_d\,\sigma^2/M$ as suggested along the way to obtain a reasonable $\mu$ \cite{KN84}. 

\begin{figure}[t!]
\centerline{\includegraphics[width=0.56\textwidth]{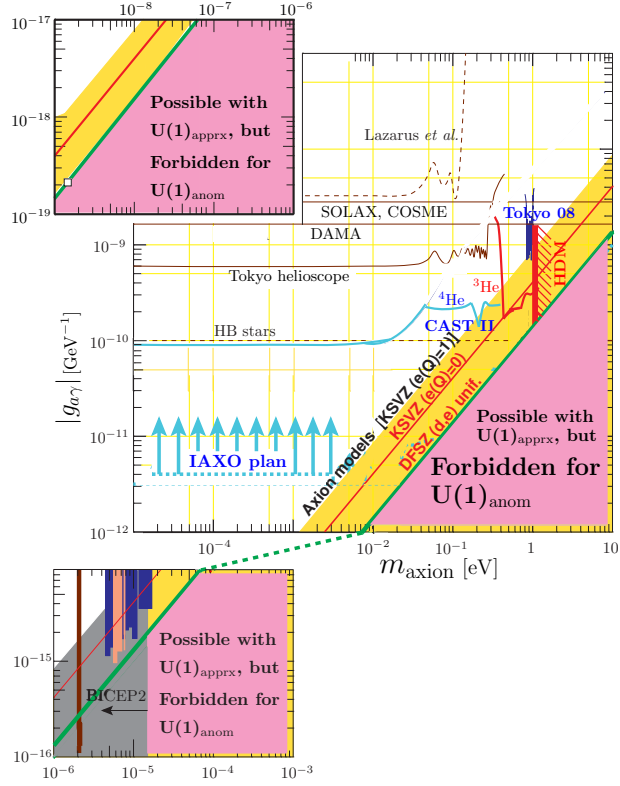}
}
\caption{Axion search bounds and some model lines.}\label{Fig:Data}
\end{figure}
\begin{wrapfigure}{r}{0.5\textwidth} 
{\includegraphics{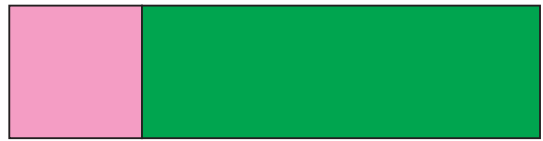}
}
\caption{Global symmetry breaking by anomalies. If an axion is present for a non-Abelian gauge group $G_N$, then the vacuum angle  $\theta_N=0$ at the minimum of that axion potential.}\label{Fig:Theta}
\end{wrapfigure}

This $\mu$ term belongs to the lavender part of Fig. \ref{Fig:Discrete}\,(b). Furthermore, if the U(1) global symmetry so defined is exact, then there is no superpotential terms, \ie there is no red part above the lavender box.
Then, the situation is as shown in Fig. \ref{Fig:Theta}, and the PQ symmetry defined by 
 $H_uH_d\,\sigma^2/M$ is exact.  However, the red boxes disjoint from the horizontal green are present, \ie the anomaly terms are present, breaking this global symmetry. If this global symmetry arises from anomalous U(1) gauge symmetry from string compactification \cite{Uoneanom}, then the resulting global symmetry is free of gravity spoil and can be a good PQ symmetry for the ``invisible'' axion 
\UG. The axion-photon-photon couplings of string axions are listed in \cite{KimNam16}, and the model-independent axion point is shown in \cite{KimPLB16}, which is reproduced in Fig. \ref{Fig:Data}. The lavender part of  Fig. \ref{Fig:Data} is forbidden if the PQ global symmetry is \UG. However, if the PQ global symmetry is approximate, this lavender part is also allowed as shown in Ref. \cite{Kimjhep07}.
 
\section{The QCD axion in cosmology}

The axion solution of the strong CP problem is a cosmological solution. The vacuum of the ``invisible'' axion potential generated by the gluon anomaly is at $\bar\theta=0$ \cite{Vafa84}. If the axion vacuum starts from $a/f_a=\theta_1\ne 0$, then the vacuum oscillates and this collective motion behaves like cold dark matter(CDM) \cite{BCM14}, for which a recent calculation energy density of coherent oscillation, $\rho_a$, is given in \cite{Bae08}.

Topological defects of global \UG~produce an additional  axion energy density by the decay of string-wall system, $\rho_{st}$, for which a recent calculation for $\NDW=1$ models has been given in \cite{SekiguchiT} as  $\rho_{st}\sim O(10)\,\rho_a$.

The cosmological axion domain wall number should be 1 \cite{Sikivie82}, which is an important constraint in any axion models.
Identifying different vacua has started with Lazarides and Shafi \cite{LS82}, and the most important application obtaining $\NDW=1$ is realized in the Goldstone boson direction \cite{Choi85,KimPLB16}.

\section{Detection of axions}
Since most other participants in this conference will discuss on the possibility of detecting ``invisible'' axions, I will present here just the exclusion plot in the interaction versus axion mass plane, shown in Fig. \ref{Fig:Data}. The detection rate in the cavity experiments are calculated in axio-electrodynamics in Ref. \cite{Hong14}.

\section{Conclusion}

Here, I discussed why we need ``invisible'' axions for a solution of strong CP problem.
One prospective global symmetry toward ``invisible'' axion is the global symmetry \UG, obtained from anomalous U(1) gauge symmetry. In this case, we argued that there is no obstruction of \UG~by quantum gravity effect. We also commented some cosmological problems, and the ``invisible'' axion from \UG~has $\NDW=1$ \cite{Witten85,KimPLB16}.

\section*{Acknowledgments}
This work is supported in part by  the IBS (IBS-R017-D1-2016-a00).
 

\begin{footnotesize}

\end{footnotesize}
\end{document}